\documentclass[useAMS]{mn2e}
\usepackage{times}
\usepackage[dvips]{graphicx}

\def\ep{\varepsilon}
\def\be{\begin{equation}}
\def\ee{\end{equation}}
\def\bea{\begin{eqnarray}}
\def\eea{\end{eqnarray}}

\title[Super-Earth tidal evolution]
{Tidal decay and orbital circularization in close-in two-planet systems}
\author[A. Rodr\'iguez et al.]{A.~Rodr\'iguez,$^1$\thanks{E-mail: adrian@astro.iag.usp.br} S.~Ferraz-Mello,$^1$ T. A.~Michtchenko,$^1$ C.~Beaug\'e$^2$ and O.~Miloni$^3$\\
$^1$ Insituto de Astronomia, Geof\'isica e Ci\^encias Atmosf\'ericas, IAG-USP, Rua do Mat\~ao 1226, 05508-900 S\~ao Paulo, Brazil\\
$^2$ Observat\'orio Astron\'omico, Universidad Nacional de C\'ordoba, Laprida 854, (X5000BGR) C\'ordoba, Argentina\\
$^3$ Facultad de Ciencias Astron\'omicas y Geof\'isicas, Universidad Nacional de La Plata, Paseo del Bosque S/N B1900 FWA, La Plata, Argentina}

\date{Released 2011 Xxxxx XX}

\begin{document}

\maketitle

\begin{abstract}

The motion of two planets around a Sun-like star under the combined effects of mutual interaction and tidal dissipation is investigated. The secular behaviour of the system is analyzed using two different approaches. First, we solve the exact equations of motion through the numerical simulation of the system evolution. In addition to the orbital decay and circularization, we show that the final configuration of the system is affected by the shrink of the inner orbit. Our second approach consist in the analysis of the stationary solutions of mean equations of motion based on a Hamiltonian formalism.

We consider the case of a hot super-Earth planet with a more massive outer companion. As a real example, the CoRoT-7 system is analyzed solving the exact and mean equations of motion. The star-planet tidal interaction produces orbital decay and circularization of the orbit of CoRoT-7b. In addition, the long-term tidal evolution is such that the eccentricity of CoRoT-7c is also circularized and a pair of final circular orbits is obtained. A curve in the space of eccentricities can be constructed through the computation of stationary solutions of mean equations including dissipation. The application to CoRoT-7 system shows that the stationary curve agrees with the result of numerical simulations of exact equations.

A similar investigation performed in a super-Earth-Jupiter two-planet system shows that the doubly circular state is accelerated when there is a significant orbital migration of the inner planet, in comparison with previous results were migration is neglected.

\end{abstract}



\section{Introduction}

It is well known that close-in planets undergo tidal interactions with their host stars. The tidal effect produces orbital decay, circularization and spin-orbit synchronization of a planet orbiting a slow-rotating star (Dobbs-Dixon, Lin \& Mardling  2004; Ferraz-Mello, Rodr\'iguez \& Hussmann 2008; Jackson, Greenberg \& Barnes 2008). When the tidally affected planet has an eccentric companion, its evolution is more complicated. As we shall show in this paper for two different systems, new features appear due to the interplay of tidal effects and mutual interactions between planets.

Several works have already investigated the evolution of two-planet systems accounting for tidal dissipation (Wu \& Goldreich 2002; Mardling \& Lin 2004; Mardling 2007; Zhou \& Lin 2008, Mardling 2010, Greenberg \& Van Laerhoven 2011). Particularly in Mardling (2007), a secular model based on the Legendre expansion of the disturbing forces up to fourth order in semi-major axes ratio, $a_1/a_2$, was introduced. She has shown that the outer companion excites the eccentricity of the inner planet, accelerating its migration toward the star due to tidal dissipation. Moreover, the evolution of the system under tidal effect follows several stages, depending on the circulation or oscillation of the angle between the lines of apses. The excitation of the inner planet eccentricity reaches a quasi-equilibrium value and, in a long time-scale (which depends on physical parameters and initial orbital configuration), both eccentricities are damped to zero. If the companion planet can sustain a non-zero eccentricity of the inner planet for times of the order of the age of the system, the mechanism could explain some observed oversized planets, since the body can inflate in response to tidal heat. The companion planet hypothesis would thus explain the large size of some planets, e.g. HD 209458b, HAT-P1b and WASP-1b (see Mardling 2007 for details).

Although Mardling's model has no restrictions on the values of the planetary mass ratio, its applications have given priority to systems with more massive inner planets\footnote{Except in the case of HAT-P-13 non-coplanar system (see Mardling 2010)}. In this case, in a first approximation, we may assume that the planetary semi-major axes do not vary during the tidal evolution. This approximation may be sufficient for systems with a less massive outer planet, in which case the excitation of the inner planet eccentricity is not strong enough to induce its rapid orbital decay due to the tidal friction. As a consequence, the orbital decay of the inner planet occurs in time-scales much longer than the ages of the systems analyzed in Mardling (2007). In this scenario, analytical results have shown a good agreement with results of numerical simulations, performed through numerical integrations of the Lagrange's planetary equations.

In the present work, we use a different approach, with no restrictions on the semi-major axes variation, which allows us to investigate the tidal evolution of two-planet systems with an arbitrary mass ratio. We perform numerical simulations, using Newton's equations of the coplanar motion of two planets and including the forces due to tidal friction according to Mignard (1979). We show that the angular momentum conservation constrain the variation of the planetary orbital elements, even in the case of rapid orbital decay.

We perform an adaptation of the conservative analytical model of secular motion to the systems with slow loss of the energy, such as tidally affected pair of planets. Assuming that the planets are far enough from any mean-motion resonances, only secular terms are maintained in the disturbing function. Previous works have shown that for a given value of the total angular momentum, all possible motions of the secular system are constant-amplitude oscillations around one of the two equilibria of the secular Hamiltonian, known as Mode I and Mode II (Michtchenko \& Ferraz-Mello 2001, Michtchenko \& Malhotra 2004). We also know that the location of the stationary solutions in the phase space is uniquely defined by two parameters: the ratios of planet masses and semi-major axes.

When we consider the non-conservative secular system, in which the rate of dissipation is sufficiently slow (more precisely, it is slower than the proper frequency of the system), the parameter $a_1/a_2$ varies slightly due to the slow orbital decay of the inner planet. For all possible values of $a_1/a_2$ and the total angular momentum, we may calculate the locations of the equilibria in the phase space. A curve of stationary solutions can be constructed and then compared with the result of the numerical simulation of the system (see Michtchenko \& Rodr\'iguez 2011 for a detailed discussion).

In this paper, we restrict our investigations to systems with a hot super-Earth and a more massive outer companion. By super-Earth, we mean planets with an upper limit of $10\,m_{\oplus}$, where $m_{\oplus}$ is the mass of the Earth (Fortney, Marley \& Barnes 2007; Valencia, Sasselov \& O'Connell 2007). Up to date, twenty-four known super-Earth have been discovered with masses ranging from $1.94\,m_{\oplus}$ to $9.22\,m_{\oplus}$ and the majority of these planets evolve in multi-planet systems\footnote{http://exoplanet.eu/}. Almost half of the known super-Earth have orbital periods smaller than 5 days; so small periods indicate that the planets have experienced an orbital decay due to the tidal interaction with the star.

We apply our approach to the CoRoT-7 extrasolar system, which is composed by one super-Earth and a more massive outer companion (Queloz et al. 2009; L\'eger at al. 2009; Ferraz-Mello et al. 2011).  Previous investigations involving tidal evolution of CoRoT-7 planets include the analysis of tidal heating of CoRoT-7b and the consequences for its orbital migration. It was shown that the past history of the planet may have included a tidal heating large enough to be compared with the corresponding tidal heating of the Jovian satellite Io (Barnes et al. 2010). The coupled interaction between tides and secular evolution in the CoRoT-7 system is investigated in this paper showing that a final orbital configuration with two circular orbits is obtained in a few Myr.


This paper is organized as follow: In Section \ref{1planeta-sec} we briefly describe the tidal evolution for a single-planet system. In Section \ref{2planetas} we analyze the two-planet secular behaviour through the numerical integrations of the exact equations of motion including the tidal force (in the coplanar case). We consider two applications: the real system CoRoT-7 and an hypothetical system containing a hot super-Earth with a Jupiter companion. The angular momentum conservation of the system is discussed in Section \ref{AM-sec}, providing a simple explanation for the  damping of the outer planet eccentricity and the orbital decay of the inner planet. In Section \ref{secular-sec}, we investigate the secular dynamics of CoRoT-7 planets through the study of stationary solutions of the mean equations using the  Hamiltonian formalism. The role of stellar tides is analyzed in Section \ref{estelar-corot-sec}. Finally, Section \ref{discussion-sec} is devoted to provide a general discussion and conclusions.

\section{Tidal evolution of single-planet}\label{1planeta-sec}	

We first consider the tidal interaction between one planet and its host star. We call $m_0$, $R_0$, the mass and stellar radius, whereas $m_1$, $R_1$ are the corresponding planetary values. In the case of zero orbital inclination, the mean variations of orbital elements produced by the joint effect of both planetary and stellar tides are given by

\begin{equation}\label{adot_cumul}
<\dot{a}>\,=-\frac{4}{3}na^{-4}\hat{s}[(1+23e^2)+7e^2D]
\end{equation}
and
\begin{equation}\label{edot_cumul}
<\dot{e}>\,=-\frac{2}{3}nea^{-5}\hat{s}[9+7D],
\end{equation}
where $D\equiv\hat{p}/2\hat{s}$ and
\begin{eqnarray}\label{ps}
\hat{s}\equiv\frac{9}{4}\frac{k_0}{Q_0}\frac{m_1}{m_0}R_0^5,\quad\hat{p}\equiv\frac{9}{2}\frac{k_1}{Q_1}\frac{m_0}{m_1}R_1^5,
\end{eqnarray}
are the strengths of the stellar and planetary tides, respectively (Dobbs-Dixon et al. 2004; Ferraz-Mello et al. 2008). In above equations, $a,e$ and $n$ are semi-major axis, eccentricity and mean orbital motion, respectively. $k_i$ and $Q_i$ are the Love number and dissipation function of the deformed body, for $i=0,1$. The dissipation function is associated to the efficiency for which the energy is dissipated in the interior of the tidally deformed body. The results are valid up to second order in eccentricity, assuming a quasi-synchronous planetary rotation and a slow-rotating star (see Section \ref{corot-num-rotation-sec}). To obtain the variations produced by stellar tide, a linear model with constant time lag was assumed to relate phase lags and tidal frequencies (see Section \ref{num-sec}).

From equations (\ref{adot_cumul}) and (\ref{edot_cumul}) we obtain $\textrm{d}a/a=F(e,D)\textrm{d}e/e$, where

\begin{equation}\label{F}
F(e,D)\equiv\frac{2[1+(7D+23)e^2]}{7D+9}.
\end{equation}
After integration, we obtain

\begin{equation}\label{a-e_5}
a=a_{\textrm{\scriptsize ini}}\exp\Bigg{[}\frac{(7D+23)(e^2-e_{\textrm{\scriptsize ini}}^2)+2\log{(e/e_{\textrm{\scriptsize ini}})}}{7D+9}\Bigg{]},
\end{equation}
where the subscript ${\textrm{ini}}$ indicates initial values. It should be emphasized that, when stellar tide is neglected ($D\rightarrow\infty$) we have

\begin{equation}\label{a-e_3}
a=a_{\textrm{\scriptsize ini}}\exp(e^2-e_{\textrm{\scriptsize ini}}^2),
\end{equation}
indicating that the final position of the planet is completely determined by initial values of the elements.

It is important to note from equations (\ref{adot_cumul}) and (\ref{edot_cumul}) that the cumulative effect of tides produces orbital decay and circularization on time-scales which varies according to $k_i/Q_i$\,-\,values (see Jackson et al. 2008; Rodr\'iguez \& Ferraz-Mello 2010). Moreover, the stellar tide is the only source of orbital decay after total circularization is achieved. However, we see from equation (\ref{ps}) that the strength of the stellar tide is proportional to the planetary mass $m_1$. Hence, tides on the star due to large planets are more efficient than tides produced by small planets, in order to induce a planetary migration on time-scales comparable to the age of the system.

\section{Numerical simulations of the tidal evolution of two-planet systems}\label{2planetas}

In this section we study the dynamical evolution of a system composed by two planets orbiting a central star. The purpose is to investigate how the presence of an outer companion affects the tidal evolution of the inner planet.

\subsection{Numerical model}\label{num-sec}

\begin{figure}
\begin{center}
\includegraphics[width=0.70\columnwidth,angle=0]{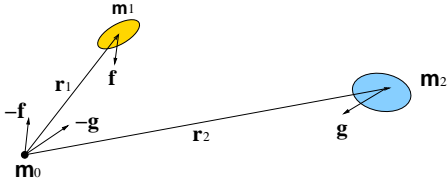}
\caption{\small Schematic illustration of tidal interaction in the system.}
\label{sistema-fig}
\end{center}
\end{figure}

We will consider systems in which the inner planet is a super-Earth. As discussed in the previous section, in this case, the contribution due to stellar tides is negligible due to the small planet mass. We assume that both planets are deformed under the tides raised by the central star. In addition to tidal force, we consider that the inner planet is also affected by the correction of Newtonian potential of the star due to general relativity. The reference frame chosen is centered in the star and the motion of the planets occurs in the reference plane (i.e coplanar motion).

According to our assumptions, the orbits evolve under the combined effects of mutual interaction and tides raised by the star. The equations of motion of the planets are written as 

\begin{eqnarray}\label{mov}
\ddot{\mathbf{r}}_1&=&-\frac{G(m_0+m_1)}{r_1^3}\mathbf{r}_1+Gm_2\Bigg{(}\frac{\mathbf{r}_2-\mathbf{r}_1}{|\mathbf{r}_2-\mathbf{r}_1|^3}-\frac{\mathbf{r}_2}{r_2^3}\Bigg{)}\\
&&+\frac{(m_0+m_1)}{m_0m_1}(\mathbf{f}+\mathbf{f}_{\textrm{\scriptsize rel}})+\frac{\mathbf{g}}{m_0},\nonumber\\
\ddot{\mathbf{r}}_2&=&-\frac{G(m_0+m_2)}{r_2^3}\mathbf{r}_2+Gm_1\Bigg{(}\frac{\mathbf{r}_1-\mathbf{r}_2}{|\mathbf{r}_1-\mathbf{r}_2|^3}-\frac{\mathbf{r}_1}{r_1^3}\Bigg{)}+\nonumber\\
&&\frac{(m_0+m_2)}{m_0m_2}\mathbf{g}+\frac{(\mathbf{f}+\mathbf{f}_{\textrm{\scriptsize rel}})}{m_0},\label{mov2}
\end{eqnarray}
where $\mathbf{f}_{\textrm{\scriptsize rel}}$ is the general relativity contribution acting on the inner planet, which is approximately given by

\begin{equation}\label{relativ}
\mathbf{f}_{\textrm{\scriptsize rel}}=\frac{Gm_1m_0}{c^2r_1^3}\left[\left(4\frac{Gm_0}{r_1}-\mathbf{v}_1^2\right)\mathbf{r}_1+4(\mathbf{r}_1\cdot\mathbf{v}_1)\mathbf{v}_1\right]
\end{equation}
where $\mathbf{v}_1=\dot{\mathbf{r}}_1$ and $c$ is the speed of light (see Beutler 2005). $\mathbf{f}$ and $\mathbf{g}$ are the tidal forces acting on the masses $m_1$ and $m_2$, respectively. We use the expression for tidal forces given by Mignard (1979):
\begin{equation}\label{mignard1-eq-pla}
\mathbf{f}=-3k_{1}\Delta t_1\frac{Gm_0^2R_1^5}{r_1^{10}}[2\mathbf{r}_1(\mathbf{r}_1\cdot\mathbf{v}_1)+r_1^2(\mathbf{r}_1\times\mathbf{\Omega}_1+\mathbf{v}_1)],
\end{equation}
\begin{equation}\label{mignard2-eq-pla}
\mathbf{g}=-3k_{2}\Delta t_2\frac{Gm_0^2R_2^5}{r_2^{10}}[2\mathbf{r}_2(\mathbf{r}_2\cdot\mathbf{v}_2)+r_2^2(\mathbf{r}_2\times\mathbf{\Omega}_2+\mathbf{v}_2)],
\end{equation}
where $\Omega_i$ is the rotation angular velocity of the $i$-planet, for $i=1,2$. It is worth noting that the Mignard's force is given by a closed formula and, therefore, is valid for any value of eccentricity\footnote{Note however that, for a more accurate description, a large number of harmonics in the expansion of the tidal potential should be considered when the star-planet distance is small enough (see Taylor and Margot 2010)}. $\Delta t_i$ is the time lag and can be interpreted as a delay in the deformation of the tidally affected body due to its internal viscosity.

The time lag is related to the dissipation function ($Q$) of the deformed planet, which is a quantity used in the study of tidal evolution of extra-solar planets. However, in many classical theories, the tidal potential appears written as a function of phase lags ($\varepsilon_j$) of tidal waves, introduced in each periodic term in order to consider the effect of viscosity (Darwin 1880; Kaula 1964; Ferraz-Mello et al. 2008). In fact, the expansion of equations (\ref{mignard1-eq-pla})-(\ref{mignard2-eq-pla}) leads to the same forces used in the Darwin's theory when the phase lags of the tide components are assumed proportional to their frequencies (the so-called linear model): $\varepsilon_j=\nu_j\Delta t$, with the same $\Delta t$ for all $j$-frequencies. For small lags, $Q$ can be associated to the phase lag of the tidal wave through $\varepsilon=1/Q$ (see Efroimsky and Lainey 2007, Efroimsky and Williams 2009 for a complete discussion). For planets in stationary or synchronous rotation, the tidal evolution is well described by the phase lag corresponding to the frequency $n$ (annual tide, see Ferraz-Mello et al. 2008). Thus, in the linear model we have $1/Q\simeq n\Delta t$ or, using a modified definition which absorbs the Love number, $Q'\equiv3Q/2k$, we obtain $k\Delta t=3/(2Q'n)$ (see Leconte et al. 2010).

\subsection{Application \#1: CoRoT-7 planets}\label{2ST-num-sec}

\begin{table}
\begin{center}
\caption{\small The adopted orbital elements and physical data of the CoRoT-7 system (L\'eger et al. 2009; Queloz et al. 2009, Ferraz-Mello et al. 2011). The value of $R_2$ was computed assuming an Earth mean density for CoRoT-7c.}\label{tabela-corot}
\begin{tabular}{|c|c|c|c|c|c|}
\hline
   Body & $m_i$  & $R_i$ & $a_{i\,{\textrm{\scriptsize current}}}$ (au)& $e_{i\,{\textrm{\scriptsize current}}}$ & $Q'_i$ \\
  \hline
  \hline
  0 & $0.93\textrm{m}_{\odot}$  & $0.87R_{\odot}$ & - & - & -\\
  \hline
  1 & $8.0m_{\oplus}$  & $1.58R_{\oplus}$ & 0.017 & 0 & 100\\
  \hline
  2 & $13.6m_{\oplus}$  & $2.39R_{\oplus}$ & 0.046 & 0 & 100\\
  \hline
\end{tabular}
\end{center}
\end{table}

We numerically investigate the recently discovered CoRoT-7 planetary system, which is composed by two short-period planets orbiting a central star. The star, CoRoT-7, is a G9V with mass $m_0=0.93\,\textrm{m}_{\odot}$, radius $R_0=0.87R_{\odot}$ and age of 1.2\,-\,2.3 Gyr (Bruntt et al. 2010). The inner planet, CoRoT-7b, and outer planet, CoRoT-7c, are assumed to have the masses given in Ferraz-Mello et al. (2011), $m_1=8.0\,m_{\oplus}$ and $m_2=13.6\,m_{\oplus}$, respectively. The radius of CoRoT-7c is not known. We adopted values of $Q'$ and mean density consistent with a rocky terrestrial planet. Their orbital periods are $P_{1\,\textrm{\scriptsize orb}}=0.854$ and $P_{2\,\textrm{\scriptsize orb}}=3.698$ days (L\'eger et al. 2009; Queloz et al. 2009). CoRoT-7b is the third exoplanet with shortest orbital period discovered up to the present date.

The adopted physical and orbital parameters of the system are listed in Table \ref{tabela-corot}. As we will shown, in the current configuration of the system, both orbits must be circular. We  emphasize that the masses of the CoRoT-7 planets satisfy the relation $m_1/m_2<1$. Choosing $Q'_1=Q'_2=100$ (a typical value for terrestrial planets) and using current values of semi-major axes, we obtain $k_{1}\Delta t_1=3$ min and $k_{2}\Delta t_2=12$ min. A scaling factor is often used in the numerical simulations in order to accelerate the process of tidal evolution. In the simulations reported in this paper, we did not use a scaling factor. We also assume that the inclinations of both orbital planes with respect to the reference plane are zero.

\begin{figure}
\begin{center}
\includegraphics[width=1\columnwidth,angle=270]{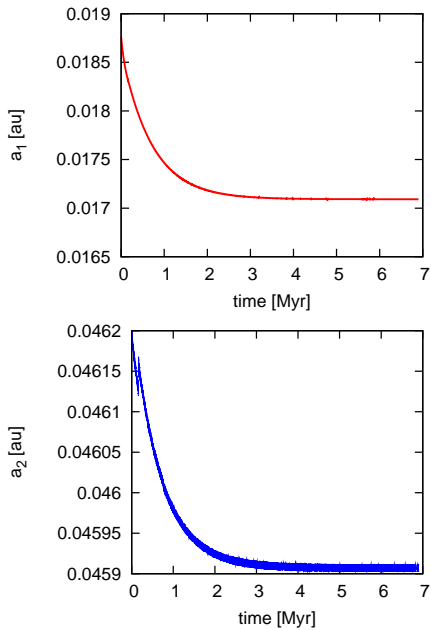}
\caption{\small Time variation of semi-major axes for the planets of CoRoT-7 system. The excitation of $e_1$ enables the orbital decay of CoRoT-7b due to tides on the planet raised by the star. CoRoT-7c also migrates due to tides but with a slower rate, resulting in a divergent migration between the planets.}
\label{corot7-fig1}
\end{center}
\end{figure}

The next step to perform the simulation is the choice of the initial configuration of the CoRoT-7 system. (It should be emphasized that, since we are studying a dissipative process, we cannot use a simple backward simulation for that sake). The choice is arbitrary, but it can be constrained by some considerations. We assume that the planets have experienced an orbital decay due to tidal effects, thus its past location must be more distant from the central star than the current one. Finally, as will be discuss in Section \ref{AM-sec}, the angular momentum of the system can be considered as invariable during the evolution of the system under the interplay of the gravitational and tidal forces. Gathering this information, we assume the following initial values for the orbital elements: $e_{1\,{\textrm{\scriptsize ini}}}=0$, $e_{2\,{\textrm{\scriptsize ini}}}=0.2$, $a_{1\,{\textrm{\scriptsize ini}}}=0.0188$ au and $a_{2\,{\textrm{\scriptsize ini}}}=0.0462$ au.

\begin{figure}
\begin{center}
\includegraphics[width=1\columnwidth,angle=270]{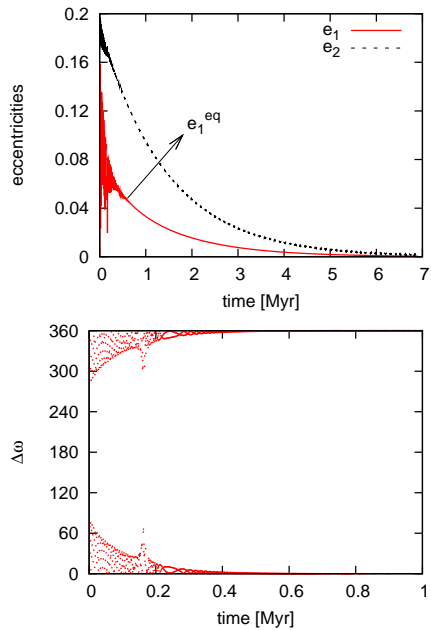}
\caption{\small Time variation of eccentricities and $\Delta\varpi$ (in degrees) for the planets of the CoRoT-7 system. The eccentricity of the initial inner circular orbit is excited by the outer companion. Tides on the inner planet circularize its orbit in approximately 7 Myr. The eccentricity of the outer orbit is also damped because of the total angular momentum conservation (see Section \ref{AM-sec}). At the end of the simulation, a state of double circularization is obtained (see text for discussion).}
\label{corot7-fig2}
\end{center}
\end{figure}

Figs. \ref{corot7-fig1} and \ref{corot7-fig2} show the result of a numerical integration of equations (\ref{mov}) and (\ref{mov2}) using the RA15 code (Everhart 1985). It is shown the orbital decay and circularization of the orbits due to tidal effects. The inner planet eccentricity initially undergoes large variations, but, after a short time of damped oscillations, it reaches a quasi-equilibrium value (Mardling 2007, see Sec. \ref{corot-num-e1-sec}). After relaxation, $e_1$ decreases smoothly to reach near-circularization in about 7 Myr. The eccentricity of the outer planet is also damped and a pair of circular orbits is obtained as final configuration of the system. 

The excitation of $e_1$ activates the orbital decay of the inner planet, which requires a non circular orbit according to equation (\ref{adot_cumul}) (with $\hat{s}=0$, because we are not considering the contribution of stellar tides in our simulations). The migration of CoRoT-7b toward the star occurs until the circularization of its orbit. Meanwhile, $a_2$ also decreases (and is only weakly affected when the 4/1 mean-motion resonance is crossed). However, when equations (\ref{mignard1-eq-pla})-(\ref{mignard2-eq-pla}) are expanded up to third order in eccentricities (see Ferraz-Mello et al. 2008), we obtain that that the ratio of forces on CoRoT-7b and CoRoT-7c is approximately 16, indicating a more intense migration of the inner planet.

The bottom panel of Fig. \ref{corot7-fig2} shows the time variation of the secular angle $\Delta\varpi\equiv\varpi_1-\varpi_2$, where $\varpi$ means longitude of pericenter. Initially, $\Delta\varpi$ oscillates around $\Delta\varpi=0^{\circ}$ with large amplitude; this stage of the evolution corresponds to the damped oscillations of the planetary eccentricities. Reaching the relaxed state, in which the system oscillates with amplitudes close to zero, the oscillation amplitude of the secular angle also tends to zero.

It is worth mentioning that, due to the divergent migration between the planets, the capture in mean motion resonance is not possible.

\subsubsection{The short-time evolution of the eccentricity of CoRoT-7b}\label{corot-num-e1-sec}

Fig. \ref{corot7-fig2} shows that, approximately at $t=0.5$ Myr, the amplitudes of oscillation of the eccentricities are damped and the variation becomes smooth. This situation is called quasi-equilibrium. Mardling (2007) derived an expression to compute the quasi-equilibrium value of the inner planet eccentricity, $e_1^{\textrm{\scriptsize eq}}$, using the Legendre expansion of the disturbing function, up to fourth order in $a_1/a_2$. It is given by
\begin{equation}\label{e1eq-mardling}
e_1^{\textrm{\scriptsize eq}}=\frac{(5/4)(a_1/a_2)e_2\ep_2^{-2}}{|1-\sqrt{a_1/a_2}(m_1/m_2)\ep_2^{-1}+\gamma\ep_2^3|},
\end{equation}
where $\ep_2\equiv\sqrt{1-e_2^2}$ and $\gamma\equiv4(n_1a_1/c)^2(m_0/m_2)(a_2/a_1)^3$.

\begin{figure}
\begin{center}
\includegraphics[width=0.50\columnwidth,angle=270]{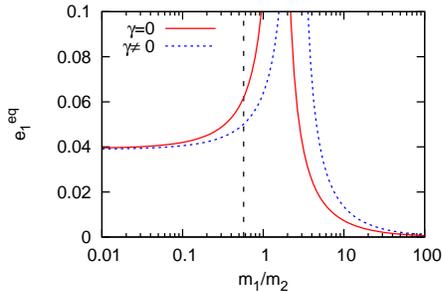}
\caption{\small Quasi-equilibrium $e_1^{\textrm{\scriptsize eq}}$ given by equation (\ref{e1eq-mardling}), as a function of the mass ratio. The vertical dashed line corresponds to the mass ratio of the CoRoT-7 system. The curves show the cases with and without general relativity.}
\label{e1eq-fig}
\end{center}
\end{figure}

Fig. \ref{e1eq-fig} shows the variation of $e_1^{\textrm{\scriptsize eq}}$ given by equation (\ref{e1eq-mardling}), for CoRoT-7, as a function of $m_1/m_2$. The relevant value of $e_2$ which should be used is the one when $e_1^{\textrm{\scriptsize eq}}$ is reached. In addition, the condition (\ref{e1eq-mardling}) was obtained under assumption that semi-major axes are constants. We show the cases with ($\gamma\neq0$) and without ($\gamma=0$) the contribution of general relativity. The inclusion of $\gamma\neq0$ reduces the value of $e_1^{\textrm{\scriptsize eq}}$. Hence, since the mean rate of orbital decay is proportional to $e_1^2$, the omission of the relativistic potential can result in an underestimation of the circularization and orbital decay time-scales. Mardling \& Lin (2004) also showed that the inclusion of the relativistic potential is very important when the planet approaches the star.

\subsubsection{The long-time evolution of the eccentricity of CoRoT-7c}\label{corot-num-e2-sec}

Our numerical simulation has shown that the eccentricity of the outer planet is strongly affected during the orbital decay (see Fig. \ref{corot7-fig1}, top panel). However, the source of $e_2$\,-\,damping is not the direct action of tides on the outer planet since, as we have already shown, the magnitude of the tidal force on CoRoT-7c is much smaller than the corresponding force on CoRoT-7b.\footnote{One can take an idea about which is the contribution of tides raised in the outer planet to the circularization process of its orbit. For that sake, we note that the ratio between the circularization time-scales associated with the direct tidal interaction with the star and the one given by equation (\ref{e2-mardling-full}) is of the order of $(a_2/a_1)^4\gg1$.}

Mardling (2007) also studied the long-term variation of the outer planet eccentricity and provided an expression to calculate the $e_2$\,-\,damping:
\begin{equation}\label{e2-mardling-full}
\dot{e}_2=-\frac{\lambda}{\tau_{\textrm{\scriptsize circ}}}\,\frac{e_2}{F(e_2)},
\end{equation}
where $\tau_{\textrm{\scriptsize circ}}$ is the circularization time-scale of the inner orbit in the single-planet case (i.e. $e_1/\dot{e}_1$, where $\dot{e}_1$ is given by equation (\ref{edot_cumul})); $\lambda\equiv(25/16)(m_1/m_2)(a_1/a_2)^{5/2}$ and $F(e_2)\equiv\ep_2^3(1-\alpha\,\ep_2^{-1}+\gamma\,\ep_2^3)^2$, with $\alpha\equiv(m_1/m_2)(a_1/a_2)^{1/2}$.

\begin{figure}
\begin{center}
\includegraphics[width=0.50\columnwidth,angle=270]{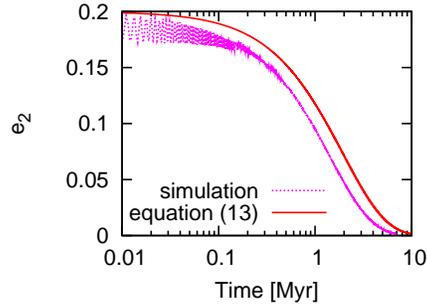}
\caption{\small Time variation of $e_2$ in the CoRoT-7 system. The comparison is done between the integration of equation (\ref{e2-mardling-full}) and the result of numerical simulation (see text for more discussion).}
\label{e2-mard-c7-fig}
\end{center}
\end{figure}

The comparison between the integration of the equation (\ref{e2-mardling-full}) and the result of the simulation of exact equations of motion (\ref{mov}) and (\ref{mov2}) is shown in Fig. \ref{e2-mard-c7-fig}. The difference between both results may arise from the fact that for obtaining equation (\ref{e2-mardling-full}) is explicitly assumed that semi-major axes are constant (see Mardling 2007)\footnote{We stress that equation (\ref{e2-mardling-full}) gives an estimate for the damping time-scale of $e_2$ (the full expression is given in equation (57) of Mardling 2007). However, it is shown in that work that the approximate solution (\ref{e2-mardling-full}) underestimates the time-scale on which $e_2$ evolves. Hence, the difference between analytical and numerical results would be even large if the full expression were used.}. In Section \ref{ST-J-num-sec}, we will show that the disagreement increases in a system for which $m_1/m_2\ll1$.

We note that, according to equation (\ref{e2-mardling-full}), the rate of $e_2$\,-\,damping decreases with $m_2/m_1$, enabling a non-circular outer orbit be maintained over long times. In addition, $e_1^{\textrm{\scriptsize eq}}$ increases with $e_2$ as equation (\ref{e1eq-mardling}) indicates. Hence, since the mean rate of orbital decay is proportional to $e_1^2$, a significant inner orbital shrinkage can occur for large $m_2/m_1$ (see next section).

\subsubsection{The rotation of CoRoT-7b and CoRoT-7c}\label{corot-num-rotation-sec}

The tidal force produces a torque that directly affects the rotation evolution of the deformed bodies, in such a way that $C_1\dot{\mathbf{\Omega}}_1=\mathbf{r}_1\times\mathbf{f}$ and $C_2\dot{\mathbf{\Omega}}_2=\mathbf{r}_2\times\mathbf{g}$. Thus,

\begin{equation}\label{mignard-torque-pla}
C_i\dot{\mathbf{\Omega}}_i=-3k_{i}\Delta t_i\frac{Gm_0^2R_i^5}{r_i^{8}}[-r_i^2\mathbf{\Omega}_i+\mathbf{r}_i\times\mathbf{v}_i],
\end{equation}
where $C_i$ is the polar moment of inertia of the $i$-planet.

\begin{figure}
\begin{center}
\includegraphics[width=0.5\columnwidth,angle=270]{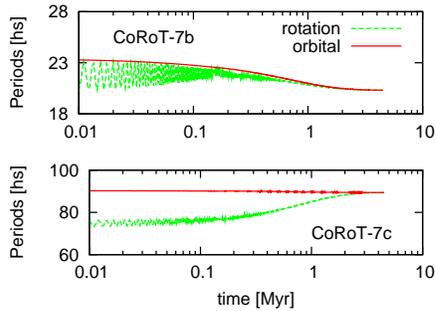}
\caption{\small Time variation of the rotation and orbital periods of CoRoT-7 planets. Synchronization is reached as the doubly circular state approaches.}
\label{rot-period-fig}
\end{center}
\end{figure}

To obtain the time evolution of the angular velocity $\Omega_i$, the above equations ($i=1,2$) were integrated simultaneously with the equations of motion (\ref{mov})\,-\,(\ref{mov2}). The result of the integration is shown in Fig. \ref{rot-period-fig}. The rotation periods, defined as $P_{i\,\textrm{\scriptsize rot}}=2\pi/\Omega_i$, reach stationary values, which, up to second order in $e_i$, are given by $P_{i\,\textrm{\scriptsize rot}}=P_{i\,\textrm{\scriptsize orb}}/(1+6e_i^2)$ (Hut 1981; also see Ferraz-Mello et al. 2008; Correia, Levrard \& Laskar 2008). In the same figure, we also plot the orbital periods defined as $P_{i\,\textrm{\scriptsize orb}}=2\pi/n_i$, which  decrease as a consequence of the orbital decay. The synchronization between the orbital motions and the rotation of the planets occurs when $\Omega_i=n_i$ or $P_{i\,\textrm{\scriptsize rot}}=P_{i\,\textrm{\scriptsize orb}}$. In  Fig. \ref{rot-period-fig} we can observe that this condition is only reached after the circularization of the planet orbits.

\subsection{Application \#2: super-Earth with a giant companion}\label{ST-J-num-sec}

\begin{figure}
\begin{center}
\includegraphics[width=0.50\columnwidth,angle=270]{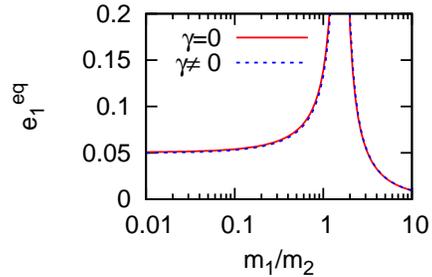}
\caption{\small Quasi-equilibrium $e_1^{\textrm{\scriptsize eq}}$ in the second application, given by equation (\ref{e1eq-mardling}), as a function of the mass ratio. The contribution of general relativity does not affect the value of $e_1^{\textrm{\scriptsize eq}}$ and thus can be neglected.}
\label{e1eq-s2-fig}
\end{center}
\end{figure}

In this section, we investigate the tidal evolution of a system in which the inner planet is much smaller than the outer one. We chose the fictitious system composed by a super-Earth and a Jupiter-like outer companion, with masses $m_1=5m_{\oplus}$ and $m_2=1m_J$, respectively ($m_J$ is the mass of Jupiter). Note that $m_1/m_2\simeq0.0157\ll1$. The initial configuration and physical parameters of this hypothetical system are listed in Table \ref{tabela-5T-1J}, in which the values of semi-major axes correspond to orbital periods of 2.92 and 11.6 days. The large orbital period of the outer planet, in addition to the large $Q$-value for Jupiter-like planets, enable us to neglect the tides on the outer companion. Moreover, as shown in Fig. \ref{e1eq-s2-fig} for the considered system, the contribution of the relativistic potential is also negligible. Here, we also assume zero inclinations. In addition, the calculation of $k_{1}\Delta t_1=3/(2Q'_1n_1)$ with $Q'_1=100$ results in $k_1\Delta t_1=10$ min.

\begin{table}
\begin{center}
\caption{\small Orbital elements and physical data corresponding to an hypothetical Sun-super-Earth-Jupiter system. The value of $R_1$ was computed assuming a terrestrial mean density.}\label{tabela-5T-1J}
\begin{tabular}{|c|c|c|c|c|c|}
\hline
   Body & $m_i$  & $R_i$ & $a_i$ (au)& $e_i$ & $Q'_i$ \\
  \hline
  \hline
  0 & $1\textrm{m}_{\odot}$  & $1R_{\odot}$ & - & - & -\\
  \hline
  1 & $5m_{\oplus}$  & $5^{1/3}R_{\oplus}$ & 0.04 & 0.1 & 100\\
  \hline
  2 & $1m_J$  & - & 0.1 & 0.1 & -\\
  \hline
\end{tabular}
\end{center}
\end{table}

\subsubsection{Short-time evolution (50 Myr)}\label{5T-1J-short-sec}

\begin{figure}
\begin{center}
\includegraphics[width=1.0\columnwidth,angle=270]{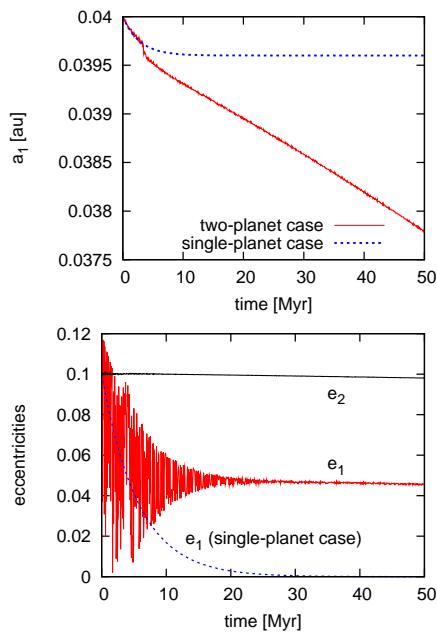}
\caption{\small Time variations of orbital elements in the case $m_1/m_2\ll1$ (c.f Table \ref{tabela-5T-1J}). The evolution in the case in which the super-Earth is the only planet is also shown. The inner orbit reaches a quasi equilibrium value in 25 Myr, retarding the orbital circularization and enhancing the orbital decay.}
\label{5T-1J-short}
\end{center}
\end{figure}

Fig. \ref{5T-1J-short} shows the time evolution of the inner planet semi-major axis (solid curve in the top panel) and two eccentricities (bottom panel) over the first 50 Myr. Due to the large mass of the outer planet, $e_2$ is almost constant over the time interval shown in the figure (see equation (\ref{e2-mardling-full})). On the contrary, $e_1$ initially suffers large oscillations which are quickly damped to the quasi-equilibrium value. On the same graph, we plot the variation of $e_1$ in the case of a single-planet system, when the interactions with one outer planet do not exist. Comparing the two cases, we note that the presence of the outer companion delays the circularization of the inner planet orbit (see Fig. \ref{5T-1J-long}). As a consequence, the presence of the second planet lead to a faster decay of the inner planet orbit. (In the single-planet case, the semi-major axis $a_1$ decreases only slightly and reaches a final constant value (0.0396 au) after total circularization)\footnote{This value can be obtained using equation (\ref{a-e_3}) with $e=0$, $e_{\textrm{\scriptsize ini}}=0.1$ and $a_{\textrm{\scriptsize ini}}=0.04$ au.}.

\subsubsection{Long-time evolution (300 Myr)}\label{5T-1J-long-sec}

\begin{figure}
\begin{center}
\includegraphics[width=1.0\columnwidth,angle=270]{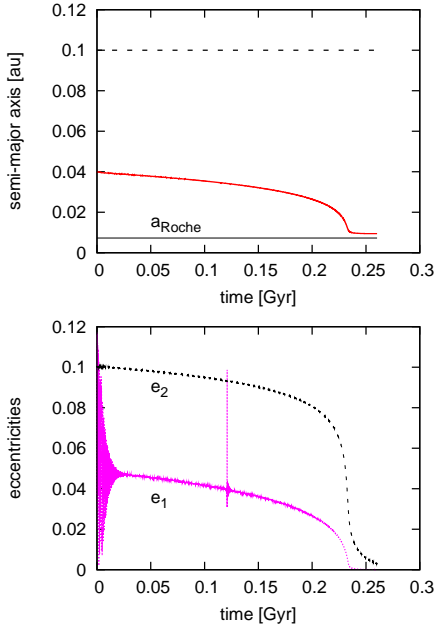}
\caption{\small Long-term evolution of semi-major axes and eccentricities in a Sun-super-Earth-Jupiter system. Due to orbital decay, the final position of the inner planet is close to the Roche limit. The value of $e_1$ is damped to zero and the outer orbit is also circularized.}
\label{5T-1J-long}
\end{center}
\end{figure}

Fig. \ref{5T-1J-long} shows the time variation of the planet semi-major axes (top panel) and the eccentricities (bottom panel) over 300 Myr. After $e_1^{\textrm{\scriptsize eq}}$ is reached, the inner planet eccentricity decreases smoothly to very small values. The outer eccentricity follows the evolution of the inner one and, at the end of simulation, about $\simeq0.25$\,Gyr, a pair of circular orbits is obtained. When the time approximates to $0.12$\,Gyr, $e_1$ is strongly excited due to the passage of the system through the 5/1 mean-motion resonance but, immediately after that, is damped to its previous value.

The orbital decay of the inner planet occurs until the total circularization of its orbit (see top panel in Fig. \ref{5T-1J-long}). Note that the final position of the super-Earth is close to the Roche limit  $a_{\textrm{\scriptsize Roche}}=(R_1/0.462)(m_0/m_1)^{1/3}$ (Faber, Rasio \& Willems 2005), which gives $a_{\textrm{\scriptsize Roche}}=0.0064$ au. The outer planet semi-major axis does not varies in the considered time interval.

\begin{figure}
\begin{center}
\includegraphics[width=0.5\columnwidth,angle=270]{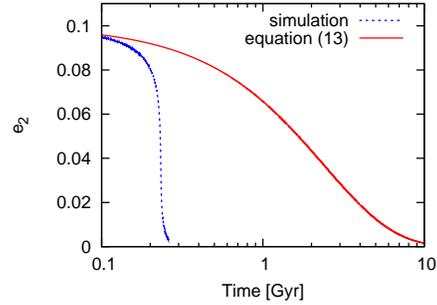}
\caption{\small Time variation of $e_2$ in a Sun-super-Earth-Jupiter system. The comparison is done between the integration of equation (\ref{e2-mardling-full}) and the result of numerical simulation (see text for more discussion).}
\label{e2-mardling2}
\end{center}
\end{figure}

In Fig. \ref{e2-mardling2} we compare the damping of $e_2$ resulting from numerical simulation (dashed line) and that obtained through the integration of equation (\ref{e2-mardling-full}) of Mardling's model (solid line). The results are very different, indicating that equation (\ref{e2-mardling-full}) is not a good approximation to compute the decreasing of $e_2$ in the case of the massive outer companion. The discrepancy can be understood taking into account that, for this particular system, there is a significant orbital decay due to tides on the inner planet, while the equation (\ref{e2-mardling-full}) was obtained assuming $\lambda$ and $\alpha$ as constants, i.e. constant semi-major axes (Mardling 2007). Thus, it is shown that the process of outer planet circularization is accelerated in the case of significant orbital decay of the inner planet, if compared with the result of equation (\ref{e2-mardling-full}).

\section{Conservation of the angular momentum}\label{AM-sec}

Many dynamical features of the planet evolution shown in the previous sections, can be explained considering the conservation of the angular momentum of the system. The total angular momentum of the three-body system in the astrocentric reference is the vectorial sum of the orbital momenta of the planets, $\mathbf{L}_i^{\textrm{\scriptsize orb}}$ ($i=1,2$), and the rotational momenta of the planets and star, $\mathbf{L}_i^{\textrm{\scriptsize rot}}$ ($i=0,1,2$).

If $\mathbf{r}_i$ is the astrocentrical position of the $i$-planet, the orbital component of the angular momentum is given by \begin{equation}\label{momtot}
\mathbf{L}_{\textrm{\scriptsize orb}}=\sum_{i=1}^{2}m_i\mathbf{r}_i\times\dot{\mathbf{r}}_i-\frac{1}{M}\sum_{i=1}^{2}m_i\mathbf{r}_i\times\sum_{i=1}^{2}m_i\dot{\mathbf{r}}_i,
\end{equation}
where $M\equiv\sum_{i=0}^{2}m_i$. Note that the second term of the equation (\ref{momtot}) appears due to the use of the astrocentric reference frame. This term is of second order in masses and, assuming $m_i\ll m_0$, can be neglected. Moreover, $\mathbf{r}_i\times\dot{\mathbf{r}}_i=\sqrt{G(m_0+m_i)a_i(1-e_i^2)}\,\hat{\mathbf{k}}$, where $\hat{\mathbf{k}}$ is a unitary vector normal to orbital planes. Hence

\begin{equation}\label{momtotredfim}
\mathbf{L}_{\textrm{\scriptsize orb}}\simeq\sum_{i=1}^2m'_i\sqrt{a_i(1-e_i^2)}\,\hat{\mathbf{k}},
\end{equation}
where $m'_i\equiv m_i\sqrt{Gm_0}$.

The rotation component of the angular momentum is given by 
$$\mathbf{L}_{i}^{\textrm{\scriptsize rot}}=\sum_{i=0}^2C_i\mathbf{\Omega}_i,$$
where $C_i$ are the moment of inertia with respect to the axes of rotation and $\mathbf{\Omega}_i$ are the angular velocities of rotation, for $i=0,1,2$. Because we account only for the planetary tides on the planets, $\mathbf{\Omega}_1$ and $\mathbf{\Omega}_2$ vary with time, while $\mathbf{\Omega}_0$ is constant.

We can show that for planets, $L_i^{\textrm{\scriptsize rot}}\ll L_i^{\textrm{\scriptsize orb}}$; thus, the contribution of the rotational component to the total angular momentum can be neglected. Indeed, using the third Kepler's law, we have $L_i^{\textrm{\scriptsize rot}}/L_i^{\textrm{\scriptsize orb}}=\xi_i(R_i/a_i)^2(\Omega_i/n_i)$, where we have introduced $C_i=\xi_im_iR_i^2$, with $0<\xi_i\le2/5$. In the case of the quasi-synchronous planet discussed in Section \ref{corot-num-rotation-sec}, when $\Omega_i\simeq n_i$, this ratio is small, of the order of $(R_i/a_i)^2$.

In contrast, for the star, the quantity $L_0^{\textrm{\scriptsize rot}}/L_i^{\textrm{\scriptsize orb}}=\xi_0(m_0/m_i)(R_0/a_i)^2(\Omega_0/n_i)$ is close to unity, when we consider close-in planets and a slow-rotating star, for which the condition $\Omega_0\ll n_i$ is satisfied. Hence, the total angular momentum of the system can be approximately written as
\begin{equation}\label{Ltot}
\mathbf{L}\simeq(m'_1\sqrt{a_1(1-e_1^2)}+m'_2\sqrt{a_2(1-e_2^2)}+C_0\Omega_0)\hat{\mathbf{k}}.
\end{equation}

Is should be noted that some quantity of angular momentum is removed due to the orbital decay of the inner planet. Thus, according to above equation, for small $e_1$, the eccentricity of the outer planet would be damped in order to preserve the angular momentum conservatation.

\subsection{The orbital decay}\label{a1-sec}

As a consequence of the total angular momentum conservation, we can obtain the minimal possible value for $a_1/a_2$ during the orbital decay, knowing that this minimum value should be achieved once orbital circularization is reached. Imposing $e_1=e_2=0$ in equation (\ref{Ltot}) we obtain

\begin{equation}\label{a2}
\frac{a_{1 \textrm{\scriptsize min}}}{a_2}=(L'-1)^2\left(\frac{m_2}{m_1}\right)^2,
\end{equation}
where $L'\equiv (L-C_0\Omega_0)/m'_2\sqrt{a_2}$ and $m'_2/m'_1\simeq m_2/m_1$. For those cases in which tides on the outer planet can be neglected, $a_2$ and thus $L'$ are constants of the problem. Applying equation (\ref{a2}) to the super-Earth-Jupiter system we obtain $a_{1 \textrm{\scriptsize min}}/a_2\simeq0.096$, reproducing the corresponding result of the numerical simulation. Hence, it is worth noting that the final position of the inner planet can be obtained from the analysis of the angular momentum conservation, as given by equation (\ref{a2}).

Equation (\ref{a2}) allows us to investigate the variation of $a_{1 \textrm{\scriptsize min}}/a_2$ for several mass ratios. However, we note that there is also an implicit dependence on $m_2/m_1$ through $L'$, which is not shown in equation (\ref{a2}). Hence, in order to express the full dependence we define two constants, namely, $\kappa_1\equiv\sqrt{(a_{1\,\textrm{\scriptsize ini}}/a_{2\,\textrm{\scriptsize ini}})(1-e_{1\,\textrm{\scriptsize \textrm{\scriptsize ini}}}^2)}$ and $\kappa_2\equiv\sqrt{1-e_{2 \,\textrm{\scriptsize ini}}^2}$. Using the definition of $L'$ and replacing it into equation (\ref{a2}), we obtain

\begin{equation}\label{a1min-full}
a_{1 \textrm{\scriptsize min}}=a_2\Bigg{[}(\kappa_2-1)\frac{m_2}{m_1}+\kappa_1\Bigg{]}^2.
\end{equation}

\begin{figure}
\begin{center}
\includegraphics[width=0.75\columnwidth,angle=270]{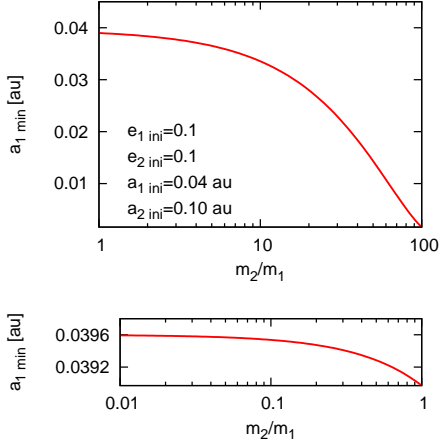}
\caption{\small Plot of equation (\ref{a1min-full}) showing the dependence of the minimum value of the inner planet semi-major axis with $m_2/m_1$. The initial elements of Table \ref{tabela-5T-1J} were used to compute $\kappa_1$ and $\kappa_2$. The case of a more massive inner planet is shown in the bottom panel.}
\label{a1min}
\end{center}
\end{figure}

Fig. \ref{a1min} shows the variation of $a_{1 \textrm{\scriptsize min}}$ with the mass ratio $m_2/m_1$ according to equation (\ref{a1min-full}), where we have used initial values of the orbital elements listed in Table \ref{tabela-5T-1J}. We see that, for fixed initial values of the elements, the inner planet will come closer to the star for high-mass companions. Therefore, as discussed in Sec. \ref{corot-num-e2-sec}, the orbital decay of the inner planet is enhanced.

For sake of completeness, we show in the bottom panel of Fig. \ref{a1min} the result for $m_2/m_1<1$, which is the case of a more massive inner planet. Note that the orbital decay is very weak, supporting the assumptions done in Mardling (2007) of constant semi-major axes in the case of slow-mass outer companions.

\section{Evolutionary routes in the phase space}\label{secular-sec}

In this section we will show that evolutionary routes of the migrating system are traced by stationary solutions of the conservative
secular problem. For this task, we explore the secular dynamics of the planar two-planet system, described in detail in Michtchenko \& Ferraz-Mello (2001) and Michtchenko \& Malhotra (2004).

In the context of hamiltonian formalism, the stationary values are defined by the secular Hamiltonian function, in which the fast-period terms, involving arguments with combinations of mean longitudes, are removed by first-order averaging. In this case, the secular problem of coplanar two-planet systems is reduced to one degree of freedom. Regarding to the angle $\Delta\varpi$, the stationary solutions are restricted to $\Delta\varpi=0^{\circ}$ and $\Delta\varpi=180^{\circ}$, which correspond to Mode I and Mode II solutions, respectively. In addition, the stationary values of eccentricities for these modes are $e_{iI}^{\ast}$ and $e_{iII}^{\ast}$, for $i=1,2$. Periodic solutions of the secular system are circulations or oscillations around Modes I or II and the stationary solution can be represented by a point in the plane of eccentricities. The position of this pont in the phase space depends on the parameters of the secular problem, namely, angular momentum and semi-major axes ratio.

To understand how the stationary solutions are affected by the introduction of dissipative forces on the system, we will assume that the dissipation is very slow. This means that the dissipation rate is much smaller than the characteristic time of the secular system, which is the proper period of the secular angle $\Delta\varpi$. Under this assumption, the motion of the pair of planets is nearly conservative and, for very short time intervals it can be described by the classical secular theory. 

For the sake of simplicity, let us consider that periodic solutions are oscillations around Mode I. When the tidal effects are taken into account, the semi-major axis of the inner planet decreases slowly. Since the semi-major axes ratio is changed, the position of the equilibrium $(e_{1I}^{\ast},e_{2I}^{\ast})$ in the phase space will be also changed, in order to preserve the total angular momentum of the system. In such a way, the system oscillates around a centre whose position in the phase space changes slowly due to dissipation.

Assuming continuous values of the inner semi-major axis, we obtain the equilibria $(e_{1I}^{\ast},e_{2I}^{\ast})$, for all possible values of $a_1$.
A curve in the space ($e_1,e_2$), which is composed by successive equilibrium points, is a locus of stationary solutions in the space of eccentricities and we refer to it as LSE. By definition, all solution in LSE curve have the same value of angular momentum.

\begin{figure}
\begin{center}
\includegraphics[width=0.5\columnwidth,angle=270]{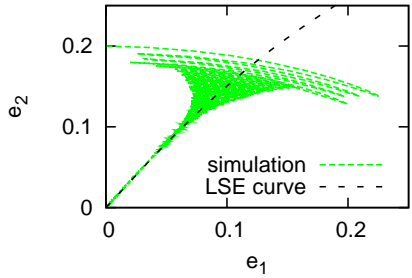}
\caption{\small Comparison between computation of LSE curve and the result of numerical simulation for CoRoT-7 system (without the contribution of the relativistic potential). After the angle $\Delta\varpi$ is close to $0^{\circ}$ (Mode I), both curves are almost coincident.}
\label{e1e2}
\end{center}
\end{figure}

Fig. \ref{e1e2} illustrates the LSE curve for Mode I in the case of the CoRoT-7 planets\footnote{We have seen in Section \ref{2ST-num-sec} that the angle $\Delta\varpi$ oscillates with almost zero amplitude around $\Delta\varpi=0^{\circ}$, indicating motion around Mode I.}, comparing with the curve resulting from numerical simulation of the system. For sake of simplicity, we do not include the relativistic potential into the analysis of mean equations. The initial conditions for the numerical integration were chosen as $\Delta\varpi=0$, $e_1=0$ and $e_2=0.2$. Initially, the eccentricities oscillate rapidly with decreasing amplitudes around an equilibrium which slides slowly along the LSE curve. When $e_1\simeq0.05$ and $e_2\simeq0.08$, the amplitude of oscillation becomes close to zero and the solution is almost coincident with the LSE curve, until the system reaches the total orbital circularization.

Detailed discussion about the secular dynamics of non-conservative two-planet systems can be found in Michtchenko \& Rodr\'iguez (2011).

\section{The role of stellar tides}\label{estelar-corot-sec}

In this section, we evaluate the time required for CoRoT-7 planets reach the Roche limit with the star. As seen in Section \ref{1planeta-sec}, in the case of small-mass planets, the contribution of tides raised by the planet on the star can be neglected if compared with the contribution of tides raised by the star on the planet. However, we know that after orbital circularization, the stellar tide is the only source of orbital decay. Thus, due to the current orbital configuration of the CoRoT-7 system, with two circular orbits, only the stellar tides could produce the orbital decay.

We define the mean lifetime of CoRoT-7b as the time for which the Roche limit is reached, starting with $a_1=a_{1\,\textrm{\scriptsize current}}=0.017$ au. According to equation (\ref{adot_cumul}) with $\hat{p}=0$, $e_1=0$ and in the limit $m_1\ll m_0$, the mean rate of semi-major axis variation is

\begin{equation}\label{adot-star-7b}
<\dot{a}_1>\,=\,-\frac{3\sqrt{Gm_0}}{a_1^{11/2}}\frac{k_{0}}{Q_{0}}\frac{m_{1}}{m_0}R_{0}^5.
\end{equation}
Equation (\ref{adot-star-7b}) can be integrated as

\begin{equation}\label{adot-star2-7b}
\int_{a_{1\,\textrm{\scriptsize current}}}^{a_\textrm{\scriptsize Roche}}da_1\,a_1^{11/2}=-{\cal{K}}\,\tau_{a_1},
\end{equation}
where ${\cal{K}}\equiv3\sqrt{Gm_0}(k_{0}/Q_{0})(m_1/m_0)R_{0}^5$, whereas $\tau_{a_1}$ is the mean lifetime of CoRoT-7b and $a_\textrm{\scriptsize Roche}=0.0049$ au ($\simeq1.2\,R_0$). Solving for $\tau_{a_1}$, we obtain

\begin{equation}\label{tau_a-7b}
\tau_{a_1}=\frac{2}{39}\frac{(a_{1\,\textrm{\scriptsize current}}^{13/2}-a_\textrm{\scriptsize Roche}^{13/2})}{\sqrt{Gm_0}R_{0}^5}\frac{Q_{0}}{k_{0}}\frac{m_0}{m_1}.
\end{equation}

\begin{figure}
\begin{center}
\includegraphics[width=0.5\columnwidth,angle=270]{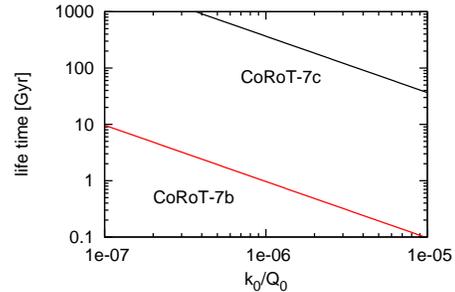}
\caption{\small The mean lifetime $\tau_{a_1}$ of CoRoT-7 planets due to stellar tides in function of the dissipation on the star. For $k_{0}/Q_0=10^{-6}$, $\tau_{a_1}=0.97$ Gyr and $\tau_{a_2}=367$ Gyr.}
\label{tau}
\end{center}
\end{figure}

Fig. \ref{tau} shows the variation of $\tau_{a_1}$ as function of $k_{0}/Q_0$ according to equation (\ref{tau_a-7b}). It is easy to see that the mean lifetime of CoRoT-7b is small for high values of $k_{0}/Q_0$. On the other hand, $\tau_{a_1}$ increases when the energy within the star is not efficiently dissipated (small $k_{0}/Q_0$ ). For a typical value of $k_{0}/Q_0=10^{-6}$, we obtain $\tau_{a_1}=0.97$ Gyr (see also Jackson, Barnes \& Greenberg  2009). If the same analysis is done for CoRoT-7c, we obtain $a_\textrm{\scriptsize Roche}=0.0062$ au ($\simeq1.5\,R_0$) and $\tau_{a_1}=367$ Gyr, indicating that CoRoT-7c should not be destroyed by tides within the lifetime of the star.

\section{Discussion and conclusions}\label{discussion-sec}

In this work we have investigated the tidal evolution of super-Earth planets with exterior companions. Previous works have already discussed that problem, giving special attention to the case of a more massive inner planet. In Mardling (2007), explicit equations were obtained giving the variation of eccentricities in the context of constant semi-major axes (see equations (\ref{e1eq-mardling}) and (\ref{e2-mardling-full})). Here, we centre the attention to the case of a more massive outer companion.

Through an approach in which the exact equations of motions are numerically solved, we have shown that the super-Earth orbital decay is accompanied by orbital circularization of both planets. In addition, we show that the rate of $e_2$-damping is small for large values of $m_2/m_1$, enabling to sustain a non-circular outer orbit for longer times. Thus, as $e_1^{\textrm{\scriptsize eq}}$ increases with $e_2$, a significant inner orbital decay can occur for large $m_2/m_1$, since its mean rate is proportional to $e_1^2$.

As a real application of tidal evolution of a super-Earth accompanied by a more massive outer planet, we investigated the CoRoT-7 two-planet system. We have shown that, starting from an initial circular orbit, $e_1$ is excited enabling the activation of the orbital decay of CoRoT-7b (which requires a non-zero value of $e_1$). The equilibrium value of $e_1$ is quickly damped by tides on the planet and a strong decreasing of $a_1$ is observed in the short-term tidal evolution.

In addition, the long-term tidal evolution has shown that the current orbital configuration of the system is achieved, including the double-circularization state of CoRoT-7 planets. After circularization, the migration toward the star continues due to the action of stellar tides. We have shown that CoRoT-7b can reach the Roche limit in approximately 1 Gyr for $k_1/Q_1=10^{-6}$ (see also Jackson et al. 2009).

The study of stationary solutions of mean equations is useful to better understand the secular dynamics in the space $e_i\cos\Delta\varpi$\,-\,$e_i\sin\Delta\varpi$. The periodic solutions around Mode I (aligned pericenters) and Mode II (anti-aligned pericenters) may be easily identified in the conservative case. In addition, the analysis can be extended to the case in which dissipation is included. It is possible to construct a curve in the space of eccentricities composed by successive stationary solutions, which are obtained one by one through the same technique used in the conservative case. We have shown that the result of numerical simulation of exact equations for CoRoT-7 planets is in good agreement with the curve of stationary solutions of mean equations of motion.

Future studies involving the dynamical evolution of a two-planet pair including tidal dissipation will help to investigate the past history of the CoRoT-7 system (i.e formation and subsequent evolution). This may bring more information about how the current orbital configuration of the system was achieved.

As an example in which $m_1/m_2\ll1$, we have also investigated the tidal evolution of a Sun-super-Earth-Jupiter-like system. The results of numerical simulation have shown that $e_2$ is not well described by integration of the simple model represented by equation (\ref{e2-mardling-full}), which describes the time variation of $e_2$ in the long-term evolution. We have shown that the doubly circular state is accelerated when there is a significant orbital migration of the inner planet, showing that the final configuration of the system is affected by the shrink of the inner orbit.

A simple analysis based on the angular momentum conservation shows that it is possible to predict the final value of the inner planet semi-major axis, provided the semi-major axis of the outer planet is fixed.

\section*{Acknowledgments}
The authors acknowledge the support of this project by CNPq, FAPESP (2009/16900-5) (Brazil), and the CAPES/SECYT joint program. The authors also gratefully acknowledge the support of the Computation Centre of the University of S\~ao Paulo (LCCA-USP). We also thank the anonymous referee for his/her stimulating revision.




\end{document}